\documentclass[a4paper]{article}
\usepackage{a4wide}

\usepackage{ifthen}
\usepackage[T1]{fontenc}
\usepackage{amsfonts}
\usepackage[dvips]{graphicx}
\usepackage[hypertex,breaklinks=true,colorlinks=false]{hyperref}
\newcommand{\pr}{Phys. Rev.}

\newcommand{\pra}{Phys. Rev. A}
\newcommand{\prb}{Phys. Rev. B}
\newcommand{\prd}{Phys. Rev. D}
\newcommand{\pre}{Phys. Rev. E}
\newcommand{\prl}{Phys. Rev. Lett.}
\newcommand{\rmp}{Rev.  Mod. Phys.}

\newcommand{\jpsj}{J. Phys. Soc. Jpn.}

\newcommand{\journaldoi}[5]{\href{http://dx.doi.org/#5}{#1\ {\bf #2}, #3 (#4)}}
\newcommand{\journal}[4]{
\ifthenelse{\equal{#1}{\pr}}{\href{http://link.aps.org/abstract/PR/v#2/e#3}{\pr~{\bf #2}, #3 (#4)}}
{\ifthenelse{\equal{#1}{\pra}}{\journaldoi{#1}{#2}{#3}{#4}{10.1103/PhysRevA.#2.#3}}
{\ifthenelse{\equal{#1}{\pre}}{\journaldoi{#1}{#2}{#3}{#4}{10.1103/PhysRevE.#2.#3}}
{\ifthenelse{\equal{#1}{\prd}}{\journaldoi{#1}{#2}{#3}{#4}{10.1103/PhysRevD.#2.#3}}
{\ifthenelse{\equal{#1}{\prl}}{\journaldoi{#1}{#2}{#3}{#4}{10.1103/PhysRevLett.#2.#3}}
{\ifthenelse{\equal{#1}{\prb}}{\journaldoi{#1}{#2}{#3}{#4}{10.1103/PhysRevB.#2.#3}}
{\ifthenelse{\equal{#1}{\rmp}}{\journaldoi{#1}{#2}{#3}{#4}{10.1103/RevModPhys.#2.#3}}
{\ifthenelse{\equal{#1}{arxiv}}{\href{http://arxiv.org/abs/#2.#3}{arXiv:#2.#3}}
{\ifthenelse{\equal{#1}{cond-mat}}{\href{http://arxiv.org/abs/cond-mat/#2}{cond-mat/#2}}
{\ifthenelse{\equal{#1}{\jpsj}}{\journaldoi{#1}{#2}{#3}{#4}{10.1143/JPSJ.#2.#3}}
{#1\ {\bf #2}, #3 (#4)}
}}}}}}}}}}

\begin{document}
\author{Grégoire Misguich\\
Institut de Physique Théorique\\
CEA, IPhT, CNRS, URA 2306 \\
F-91191 Gif-sur-Yvette, France
}

\title{Quantum spin liquids\footnote{Lectures given at the {\it Les Houches} summer school 
on ``Exact Methods in Low-dimensional Statistical Physics
and Quantum Computing'' (July 2008).}}

\date{}

\maketitle
\abstract{These notes are an introduction to a few
selected theoretical ideas in the field of quantum spin liquids: classical zero modes and breakdown of the $1/S$
expansion, the Lieb-Schultz-Mattis-Hastings theorem and Oshikawa's argument, the short-ranged
resonating valence-bond picture, large-$N$ limit (Schwinger bosons) and $\mathbb{Z}_2$ gauge theory.
}

\tableofcontents

\section{Introduction: band and Mott insulators}

Depending  on  the
context (experiments, theory, simulations,...),  ``Quantum spin liquid''
is sometimes used with rather different meanings. But let us start with a first simple
definition: the ground state
of a lattice quantum spin model is said to be a quantum spin liquid (QSL)
if it spontaneously breaks {\it no} symmetry.
According to this first definition, a QSL is realized if the spins fail
to develop any kind of long range order
at zero temperature ($T=0$) (hence the word ``liquid'', as opposed to solids which
are ordered and break some symmetries).
Of course, this first definition raises a number of questions:
Does this define  new distinct states of matter ? Do QSL have some interesting properties ?
Are there some experimental examples ?
To answer these questions, it is useful to go back to the origin of magnetism in insulators.

Generally speaking, there are two kinds of insulators: {\it band} insulators, and {\it Mott} insulators.
The first ones can be qualitatively understood from the limit of non-interacting (or weakly interacting) electrons.
Consider for instance a periodic lattice\footnote{We use a tight binding model where the solid is modeled by one state per site, neglecting  (or, more precisely, integrated out) filled orbitals or high energy empty states.} with an {\it even} number $n$ of sites per unit cell, with an average electron density of one electron per site (so-called half filling).
The Hamiltonian describing how the electrons hop from sites to sites looks like $H_K=-t \sum_{\langle i,j\rangle,\sigma=\uparrow,\downarrow}\left( c^\dagger_{i\sigma}c_{i\sigma}+H.c\right)$, where
only first neighbor hopping is considered for simplicity. $H$ can be diagonalized in Fourier space
and gives $n$ dispersing bands. The ground state is just the Fermi sea obtained by filling the lowest energy states.
Since the density is one electron per site, the $n/2$ lowest energy bands are completely filled
(one up and one down electron for in each single particle state). Assuming that the 
band $n/2+1$ is separated by a gap $\Delta$ in energy from the $n/2$ lower bands, all the excitations are gapped and,
at temperatures smaller than the gap, there is no charge carrier to carry an electric current. This is the well known picture for a band insulator: there are no low energy charge degrees freedom, no magnetic (spin) degrees of freedom, the ground state (Fermi sea) is unique and breaks no symmetry. To get some interesting QSL, we should instead look at Mott insulators. There, the number of sites
per unit cell is {\it odd} and the non-interacting limit is unable to give the correct insulating behavior
(at least one band is partially filled, hence with low energy charge excitations). It is more useful to look at the system
in the opposite limit of very large electron-electron repulsion, as with the large $U$  limit of the Hubbard model:
$H=H_K+ U\sum_i, c^\dagger_{i\uparrow}c_{i\uparrow}c^\dagger_{i\downarrow}c_{i\downarrow}$. At
$U=\infty$ and $t=0$ (still at half filling), the ground state is highly degenerate ($=2^V$, where $V$ is the total number of sites) since any state with one electron per site is a ground state, whatever the spins orientations. To describe how
this degeneracy is lifted at weak but finite $t/U$, a second order perturbation has to be computed.\footnote{At first order in $t$, a single electron hopping inevitably leads to a doubly occupied site.} The result is an effective Hamiltonian
acting in the subspace spin configurations, and takes the form of a quantum spin-$\frac{1}{2}$ Heisenberg model:
\begin{equation}
	H=\frac{1}{2}\sum_{ij} J_{ij} \vec S_i \cdot \vec S_j 
	\label{eq:Heisenberg}
\end{equation}
where $J_{ij}=t_{ij}^2/U$ involves the hopping amplitude $t_{ij}$ between sites $i$ and $j$ and
measures the strength of the antiferromagnetic (AF) interaction between the (electron) spins $\vec S_i$
and $\vec S_j$.\footnote{In real materials, there are often tens or hundreds of electron per unit cell, several ions and
many atomic orbitals. Although the description of the magnetic properties in terms of lattice spin models if often very accurate, the spin-spin interactions is often more complicated than this antiferromagnetic Heisenberg model.
It is quite frequent that some interactions violate the $SU(2)$ symmetry of the Heisenberg model, due to spin-orbit couplings in a crystalline environment. In these notes, we focus on models with an $SU(2)$ symmetry.}

Although the model of Eq.~\ref{eq:Heisenberg} is in general a  complicated quantum many body problem with very few exact results,\footnote{In these notes, we focus here on dimension $D>1$, but much more is known about one-dimensional (1D) spin chains.} its ground state and low energy properties are qualitatively well  understood in many cases. In particular, the ground
state can be {\it antiferromagnetically ordered} (also called N\'eel state). Such state can be approached from a semi classical point of view described in Sec. \ref{sec:SW}: the spins point well defined directions 
and form a regular structure. Most of the Mott insulators studied experimentally  belong to this family.
The simplest example is the nearest neighbor Heisenberg model
on bipartite lattices such as the square, cubic or hexagonal lattices. There,  on average, all
the spins of the sublattice $A$ point in direction $+\vec S_0$ (spontaneous symmetry breaking of the $SU(2)$ rotation symmetry), and all the spins of sublattice $B$ point in direction $-\vec S_0$.
The difference with a classical spin configuration is that the
magnetization of one sublattice (it is the order parameter for a Néel state)
is reduced by the quantum zero-point fluctuations of the spins, even at $T=0$. Such ordered states are not QSL (they might instead be called spin ``solids'') since they break the rotation symmetry. 

The main question addressed in these notes is the fate of the ground state of Eq.~\ref{eq:Heisenberg} when the lattice and the interactions $J_{ij}$ are such that the spins {\it fail} to develop any such Néel ordered state.
A state without any order is not necessarily interesting from a theoretical point
of view. For instance, a spin system at very high temperature
is completely disordered and does not have any rich structure. As we will see,
the situation in Mott insulators at $T=0$ is completely different.
A first hint that Mott QSL host some interesting topological properties
will be discussed in Sec.~\ref{sec:LSMH} (Lieb-Shultz-Mattis \cite{lsm61} Hastings \cite{hastings04} theorem).
A concrete (but qualitative) picture for QSL wave functions is given
in Sec.~\ref{sec:RVB}, in terms of short range valence bond configurations
and deconfined spinons (magnetic excitations carrying a spin $\frac{1}{2}$).
Finally, Sec.~\ref{sec:SB}, presents a formalism which puts some of the ideas above on firmer grounds.
It is based on a large-$N$ generalization of the Heisenberg models ($SU(2)\to Sp(N)$) which allow to describe some
gapped QSL and to establish a connection by topologically ordered state of matter, such as
the ground state of Kiatev's toric code \cite{kitaev97}.

\section{Some materials without magnetic order at $T=0$}

There are many magnetic insulators that do order at $T=0$.\footnote{Some order
at a temperature with is very small compared the typical energy scale
of the Heisenberg spin-spin interactions. This is often due to perturbations that are not included
in the simplest Heisenberg model description.
}
For instance, the magnetic, properties of many compounds are described by 1D spin chains of spin ladder Hamiltonians. Thanks to the Mermin-Wagner theorem and the reduced dimensionality, these system cannot develop long range spin-spin
correlations, even at $T=0$.\footnote{Due to some residual 3D couplings, there can be a finite
temperature phase transition to an ordered state at very low temperature.} They certainly deserves to be called QSL
and represent a very rich field of activity. In these notes we will instead focus on QSL in $D>1$ systems, where our present
understanding is less complete.

\begin{figure}
\begin{center}
\includegraphics[height=4cm]{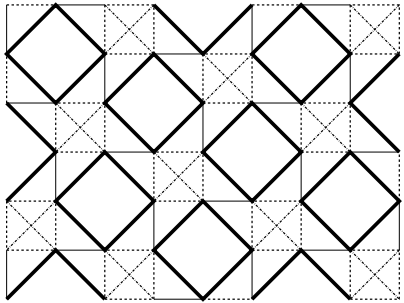}
\caption{Depleted square lattice model for the magnetic properties of CaV$_4$O$_9$.
The different exchange energies are shown by different types of line. The strongest
$J$ correspond to the fat lines forming the large tilted square plaquettes.
}\label{fig:cav4o9}
\end{center}
\end{figure}

CaV$_4$O$_9$ is the first Heisenberg  system in $D>1$ where the magnetic excitations were experimentally shown to be
gapped, in 1995 \cite{taniguchi95}.
This compound can be modeled by an antiferromagnetic spin-$\frac{1}{2}$ Heisenberg model on
a {\it depleted} square lattice where one site out of five is missing
(Fig.~\ref{fig:cav4o9}). The remaining sites correspond to the locations of the Vanadium ions, which carry the magnetically active electrons (spins). The magnetic interactions $J_{ij}$ turned out to be significant not only between
nearest neighbors, but also between second nearest neighbors (the electron hops through oxygen orbitals, which have a complex geometry).
Through magnetic susceptibility measurements, it was shown that the ground state is a rotationally
invariant spin singlet, thus excluding any Néel ordering.
This QSL behavior can be understood by taking a limit where only the strongest $J_{ij}$ are kept, and the other are set to zero. It turns out these strongest couplings are between second-nearest neighbors, and form a set of {\it decoupled}
four-site plaquettes (of area $\sqrt{2}\times\sqrt{2}$ and surrounding a missing site). Since the ground state of such a four-site Heisenberg cluster is a unique singlet $S=0$ state, separated by a gap from other states, the model is
trivially a gapped and without any broken symmetry  in this limit. But this is not the kind of QSL we want to focus on here,
since it can be adiabatically transformed into a band insulator. Switching off the electron-electron interactions
would make the system metallic, but one can proceed in a different way. Starting with realistic values
of the $J_{ij}$, the inter plaquette couplings are gradually turned off. Doing so, one can check (numerically for instance) that the spin gap does not close and no (quantum) phase transition in encountered. Then, in this systems of decoupled four-electron cluster, the Hubbard repulsion $U$ can be switched to zero, without causing any phase transition. The final model is evidently a band insulator and smoothly connected to the initial Heisenberg model.

Since then, numerous 2D and 3D (Heisenberg) magnetic systems with an {\it even} number of spin-$\frac{1}{2}$ per unit cell have been found  to be gapped. To our knowledge, their ground state can be qualitatively
understood from a limit of weakly coupled clusters in all cases and can therefore be ``classified'' as band insulators (as CaV$_4$O$_9$ above). Some of them can be very interesting for different reasons,\footnote{For instance: TlCuCl$_3$ \cite{tlcucl3} is coupled dimer system with a Bose-Einstein condensation of magnetic excitations in presence of an external magnetic field, and SrCu$_3$(BO$_3$)$_2$ \cite{k99} has a magnetization curve with
quantized plateaus.} but their ground states are not fundamentally new states of matter.

In the recent years, experimentalists have also uncovered a number of materials
which are well described by 2D Heisenberg models with an {\it odd} number of spin-$\frac{1}{2}$ per crystal unit cell,\footnote{In such case, the absence of long range order cannot be attributed to some band insulator physics.}
and which do not develop any Néel order when $T\to0$. Some examples are the Herbertsmithite (ZnCu$_3$(OH)$_6$Cl$_2$)\footnote{Although
the spin-spin interaction strength is of the order of $J\sim 200$~K, no order has be found down to 50mK.} \cite{herbert}
and Volborthite (Cu$_3$V$_2$O$_7$(OH)$_2$~$\cdot$2H$_2$O)\cite{volbo} minerals (both with a kagome lattice geometry),
triangular based organics materials \cite{shimizu03,itou08},
or triangular atomic layers of He$^3$ adsorbed onto graphite \cite{masutomi04} (there the spin is
not electronic, but nuclear).
It turns out that all these systems seem to have {\it gapless} magnetic excitations
and a complete theoretical understanding of these system is still lacking.
The present theories for gapless QSL 
are rather elaborate \cite{gaplessQSL} and many questions remain open (stability, nature of the excitations,
correlation exponents, etc.).
However, as we will see, {\it gapped} QSL are  simplest from a theoretical point of view.
Intriguingly, to our knowledge,
no gapped QSL has been discovered so far in nature, although many spin models do have gapped QSL ground states.

\section{Spin wave theory, zero modes and breakdown of the $1/S$ expansion}
\label{sec:SW}

To understand why an AF Heisenberg spin model can {\it fail} to order at zero temperature,
is is useful to briefly review the standard approach to Néel phases: the semi-classical
$1/S$ spin-wave expansion \cite{anderson52}. This approach i) starts from a classical spin configuration which
minimizes the Heisenberg interaction ii) assumes that the quantum deviations from this
ordered direction are {\it small} iii) treats this deviations as collection of harmonic oscillators
(the leading term in a $1/S$ expansion). In this approximation the Hamiltonian is written
using boson creation and annihilation operators, is quadratic, and can be diagonalized by a
Bogoliubov transformation. One can then check {\it a posteriori} if the spin deviations are indeed small.
If it is not the case, we have a strong indication that the magnetic long range order is in fact ``destroyed''
by the quantum fluctuations, thus opening a route for a QSL ground state.
\subsection{Holstein-Primakoff representation}
The starting point is the  representation of the spin operators using Holstein-Primakoff \cite{hp40} bosons
\begin{eqnarray}
	S^z_i &=& S - a^\dagger_i a_i \;,\; S^+=\sqrt{2S-a^\dagger_i a_i} \;\;a_i
	\;,\; S^-=a^\dagger_i \sqrt{2S-a^\dagger_i a_i},
	\label{eq:HP}
\end{eqnarray}
from which
on can check that the  commutation relations $[S^\alpha_i,S^\beta_i]=i\epsilon^{\alpha\beta\delta}S^\delta_i$
and $\vec S^2_i=S(S+1)$ are satisfied (using $[a_i,a_i^\dagger]=1$).

Let $\left\{\vec z_i\right\}$ be a classical ground state of Eq.~\ref{eq:Heisenberg},
minimizing $E=\frac{1}{2}\sum_{ij}J_{ij} \vec z_i\cdot\vec z_j$
with $\vec z_i^2=1$. These directions can be used as local quantization axes:  we use Eq.~\ref{eq:HP} in a
local (orthogonal) frame $(\vec x_i,\vec y_i,\vec z_i=\vec x_i \wedge  \vec y_i)$ adapted to the classical ground state.
Under the assumption that $\vec S_i$ shows small deviations from
the classical vector $S\vec z_i$, the typical number $\langle a^\dagger a\rangle$ of Holstein-Primakoff bosons
should be small compared to $S$. We can therefore simplify $S^+$ (and $S^-$) in Eq.~\ref{eq:HP} by keeping only $\sqrt{2S}$ in the square roots, to obtain \cite{anderson52}
\begin{eqnarray}
	\vec S_i \simeq \left((S+\frac{1}{2})-\vec\pi_i^2\right)\vec z_i
	+\sqrt{2S}\vec \pi_i
	\label{eq:Spi}
\end{eqnarray}
where
\begin{eqnarray}
\vec\pi_i&=&\frac{1}{2}(a_i+a_i^\dagger)\vec x_i + \frac{1}{2i}(a_i-a_i^\dagger)\vec y_i\\
\vec\pi_i^2&=&a_i^\dagger a_i+\frac{1}{2}\\
{\rm and}\; \vec z_i\cdot \vec \pi_i&=&0.
\end{eqnarray}
Replacing Eq.~\ref{eq:Spi} in the Hamiltonian
gives
\begin{eqnarray}
	H=\frac{1}{2}(S+\frac{1}{2})^2\sum_{ij}J_{ij}\; \vec z_i\cdot\vec z_j
	+ S\sum_{ij}J_{ij}\; \vec \pi_i\cdot\vec \pi_j \nonumber\\
	-\frac{1}{2}S \sum_{ij}J_{ij}\; \left( \vec\pi_i^2  +\vec\pi_j^2\right)\vec z_i\cdot\vec z_j
	+\mathcal{O}(S^0).
	\label{eq:Hpi2}
\end{eqnarray}
The first term is a constant, proportional to the classical energy $E_0$. The two other terms, proportional
to $S$, are quadratic in the boson operators and describe the spin fluctuations as a set of coupled harmonic oscillators.\footnote{Due to the fact that $\left\{\vec z_i\right\}$ minimizes the classical energy,
$\sum_j J_{ij} \vec z_j$ is perpendicular to $\vec z_i$ and thus orthogonal to $\vec \pi_i$, and there
is no term {\it  linear} in $\vec \pi$.}
The positions
$q_i=\frac{1}{\sqrt{2}}(a_i+a^\dagger_i)$
and momenta $p_i=\frac{1}{\sqrt{2}i}(a_i-a^\dagger_i)$
operators of these oscillators
can be conveniently grouped into a column vector of size $2N$ ($N$ is the total number of spins):
\begin{equation}
 {\bf V}=\left[
	\begin{array}{c} q_1\\ \colon \\q_N\\ p_1 \\ \colon \\p_N\end{array}
	\right] 
\end{equation}
so that $H$ becomes
\begin{eqnarray}
	 H=(S+\frac{1}{2})^2 E_0+\frac{S}{2} {\bf V}^t \mathcal{M} {\bf V},
	\label{eq:HM}
\end{eqnarray}
where $\mathcal{M}$ is a $2N\times2N$ matrix  given by
\begin{equation}
 \mathcal{M}=\left[
	\begin{array}{cc}
	 J^{xx}-J^{zz} & J^{xy} \\
	(J^{xy})^t & J^{yy}-J^{zz}
	\end{array}
\right]
\end{equation}
and the $N\times N$ matrices $J^{xx}$, $J^{yy}$, $J^{xy}$ and $J^{zz}$
are defined by:
\begin{eqnarray}
 J^{xx}_{ij}=J_{ij}\;\vec x_i \cdot \vec x_j \;\;,\;\; J^{yy}_{ij}=J_{ij}\;\vec y_i \cdot \vec y_j \;\;,\;\; J^{xy}_{ij}=J_{ij}\;\vec x_i \cdot \vec y_j \\
{\rm and} \;\;\; J^{zz}_{ij}=\delta_{ij} \sum_k\; J_{ik}\vec z_i\cdot \vec z_k.
\end{eqnarray}
\subsection{Bogoliubov transformation}
Diagonalizing $H$ amounts to  find bosonic creation operators $b_\alpha^\dagger$ and corresponding
energies $\omega_\alpha \geq 0$ such that
$H=\sum_\alpha \omega_\alpha \left( b_\alpha^\dagger b_\alpha+\frac{1}{2}\right)$ (up to a constant).
A necessary condition is that the operator $b_\alpha^\dagger$ and
$b_\alpha$ are ``eigenoperators'' of the commutation with $H$, for the eigenvalues
$\omega_\alpha$ and $-\omega_\alpha$ respectively:
$\left[H,b_\alpha^\dagger\right]=\omega_\alpha b_\alpha^\dagger$
and $\left[H,b_\alpha\right]=-\omega_\alpha b_\alpha$. We thus seek the eigenvectors
of the action of $\left[H,\bullet\right]$ in the space of linear combinations of $q_i$ and $p_j$.
The commutators of $H$ (Eq.~\ref{eq:HM}) with the operators $q$ and $p$ are simple to obtain
using $[q_i,q_j]=[p_i,p_j]=0$ and $[q_i,p_j]=i\delta_{ij}$. 
For an arbitrary linear combinations of the $q_i$ qnd $p_i$ parametrized
by the complex numbers $x_1,\cdots,x_{2N}$ the result is
\begin{eqnarray}
	\left[H,x_1 q_1+x_N q_N + x_{N+1}p_1+\cdots x_{2N}p_N\right] \nonumber\\
	= y_1 q_1+y_N q_N + y_{N+1}p_1+\cdots y_{2N}p_N
\end{eqnarray}
with the coefficients $y_1,\cdots,y_{2N}$ given by
\begin{eqnarray}
	\left[\begin{array}{c}y_1 \\ \vdots \\y_{2N}\end{array}\right]=i S \; \mathcal{M} \left[
		\begin{array}{ccc}
		 0 & &{\bf 1} \\
		\\
		 -{\bf 1} & & 0
		\end{array}
	\right]
	\left[\begin{array}{c}x_1 \\ \vdots \\x_{2N}\end{array}\right]
\end{eqnarray}
where ${\bf 1}$ is the $N\times N$ identity matrix.
So, finding the operators $b_\alpha^\dagger$ (spin-wave creation operators)
amounts to find the eigenvectors of the ``commutation matrix''
$\mathcal{C}=i \mathcal{M} \left[\begin{array}{cc}0&{\bf 1}\\-{\bf 1}&0\end{array}\right]$.

But $\mathcal{C}$ is not symmetric
and cannot always be fully diagonalized (contrary to $\mathcal{M}$).
It can be shown that if all the eigenvalues of $\mathcal{M}$ were {\it strictly}
positive,
$\mathcal{C}$ could be diagonalized, its eigenvalues would be real and come in pairs
$-\omega$,$\omega$.\footnote{Let $P$ be an orthogonal matrix
which diagonalizes symmetric $M$: $M=P^{-1}\lambda P$, where $\lambda$ is a diagonal matrix
and $P P^t=1$.
If the eigenvalues of $M$ are {\it strictly} positive, $K=P^{-1}\sqrt{\lambda}$ is invertible and $M=K K^t$.
We write $C=iS\;K K^t\;\sigma$, where
$\sigma=
\left[
		\begin{array}{cc}
		 0 &{\bf 1} \\
		 -{\bf 1} & 0
		\end{array}
	\right]$. Then, $\tilde C=K^{-1} C K =i S\;K^t\;\sigma K$ is Hermitian (since $\sigma$ is real antisymmetric, and K is real). $\tilde C$ can therefore be diagonalized and its spectrum is real.
Since $C$ and $\tilde C$ have the same spectrum, $C$ can also be diagonalized and
has real eigenvalues. Finally, we use $C^t=-C$. Since $C$ and $C^t$ should have the same spectrum,
the eigenvalues of $C$ go in pairs $-\omega,\omega$. 
}

However, $\mathcal{M}$ does have some zero eigenvalues.
The matrix $\mathcal{M}$ is not specific to the quantum spin problem. The quadratic form describing the
classical energy variation for a small perturbation around the chosen classical ground state
$\left\{\vec z_i\right\}$, is described by {the same matrix} $\mathcal{M}$.\footnote{The Eqs.~\ref{eq:Spi} and \ref{eq:Hpi2} also hold if $\vec\pi_i$ is a classical spin deviation of length $\vec \pi_i^2\ll1$.} In particular, if the classical ground state admits some zero energy (infinitesimal) spin rotations, $\mathcal{M}$ posses some eigenvector for the eigenvalue 0. Because global rotations should not change the energy, $\mathcal{M}$ has at least two zero eigenvalues.
Still, these global rotations do not cause difficulties
in diagonalizing the spin-wave Hamiltonian, they just correspond to some $\omega_\alpha=0$
(the associated collective coordinate $Q$ {\it and} conjugate momentum $P$ simply do not appear in $H$).

\subsection{Zero modes on the kagome lattice}

\begin{figure}
\begin{center}
 \includegraphics[height=5cm]{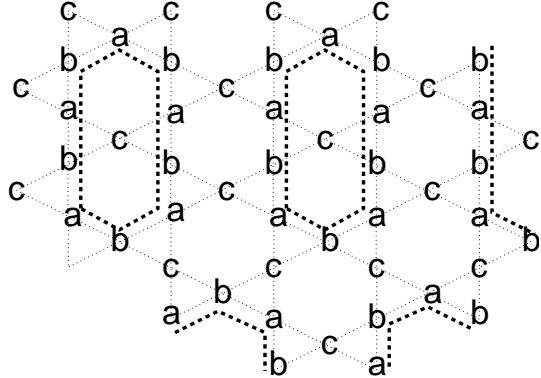}
\caption{Classical planar ground state on the kagome lattice. The loops
where the spins alternate between the $\vec a$ and $\vec b$ directions are marked with dashed
lines, they host independent zero modes (by rotation around the $\vec c$ axis).}
\label{fig:kagomeABC}
\end{center}
\end{figure}

However, some Heisenberg models admit classical zero modes (hence zero eigenvalues in $\mathcal{M}$) which do
not correspond to global rotations.
As an example, consider the Heisenberg model on the kagome lattice \cite{chs92}  (for another classic example,
the $J_1$-$J_2$ model on the square lattice, see \cite{cd88}). Any classical spin configuration
such that the sum $\vec z_i + \vec z_j + \vec z_k$ vanishes on each triangle $(ijk)$ minimizes the classical energy. Among the numerous ways to achieve these conditions, are the {\it planar} ground states, where all
the spins lie in the same plane. In such a state, the spins take only three possible directions, $\vec a$, $\vec b$ and $\vec c$ at 120 degrees from each other. On the kagome lattice, there is an exponential number of ways to assign these three orientations such that the same letter is never found twice on the same triangle
(three-coloring problem, see Fig.~\ref{fig:kagomeABC}). Now, choose one of these ``$abc$'' states, and find a closed loop of the type $ababab\cdots$. Because of the three-coloring rule, the spins which are neighbors of this loop  all point in the $\vec c$ direction. Now, we can rotate the spins of the loop about the $\vec c$ axis by any angle.
This transforms the planar ground state into another (non planar) ground state, without any energy cost.
So, for a generic planar ground state, we get as many zero modes (in $\mathcal{M}$) as closed loops
with two alternating ``colors''. This number typically grows like the number of sites in the system.

What are the consequences of such classical zero modes for  the quantum problem ?
As explained previously, the operators describing the two transverse directions along which the spins can
deviate from the $\vec z_i$ axis obey the same commutation rules (at leading order in the $1/S$ expansion) as  the
position $q$ and momentum $p$ of an harmonic oscillator. In the case of the kagome loop modes
discussed above, the energy is zero in one direction (rotation about the $\vec c$ direction), and quadratic in
the other direction. Using the associated collective coordinate $P$ and $Q$, we expect the Hamiltonian
to be proportional to
$H=\frac{1}{2}(P^2+\omega^2 Q^2)$ with $\omega=0$, since there is no classical energy cost for spin deviation in the direction $Q$.
The corresponding commutation matrix is $C=i\left[\begin{array}{cc}0&\omega^2\\-1 &0\end{array}\right]$
and cannot be diagonalized when $\omega=0$, as anticipated.
In general, each such local zero mode will lead to an irreducible $2\times2$ Jordan block of this kind.\footnote{The general theory for possible Jordan forms of $\mathcal{C}$ (size and nature of the irreducible blocks)
is in fact a result of {\it classical mechanics}, found by Williamson and exposed in \cite{arnold}.}
The ground state $|0\rangle$ of the oscillator is
simple to obtain and corresponds to a zero point motion of the coordinate $Q$ which {\it diverges}
when $\omega\to 0$ (no restoring force, like for a free particle) :
$\langle 0|Q^2|0\rangle=\frac{1}{2\omega}$.

As long as the the number of such zero modes is {\it finite} in the thermodynamic limit (this is the case
when the classical ground state has no special degeneracy, beyond those implied by global rotations),
the divergences above have a zero measure and do not cause divergences in the number of bosons $\langle 0|a^\dagger_i a_i|0\rangle$,\footnote{
$\langle 0|a^\dagger_i a_i|0\rangle=\frac{1}{2}\langle 0|p_i^2+q_i^2-1|0\rangle$ can be computed by expressing
$q_i$ and $p_i$  in terms of $b^\dagger_\alpha$ and $b_\alpha$,
or in terms of the new position and momenta $Q_\alpha=\frac{1}{\sqrt 2}(b_\alpha+b^\dagger_\alpha)$
and $P_\alpha=\frac{1}{\sqrt 2 i}(b_\alpha-b^\dagger_\alpha)$
Concentrating on the term $\langle 0|q_i^2|0\rangle$, 
$q_i$ is a linear combination of the type
$q_i=\sum_{\alpha=1}^N u_\alpha^i Q_\alpha + \sum_{\beta=1}^N v_\beta^i P_\beta$,
($u$ and $v$ are related to the eigenvectors of $\mathcal{C}$).
From the fact that $|0\rangle$ is the vacuum of the  $b_\alpha$ bosons,
we have $\langle 0|P_i P_j |0\rangle=\langle 0|Q_i Q_j |0\rangle$ if $i\ne j$,
and $\langle 0|P_i Q_j + Q_jP_i|0\rangle=0$ $\forall i,j$.
Then the square of the  spin deviation at site $i$ (here the $\vec x_i$ component) is a linear combination
of the zero point fluctuations of the normal harmonic oscillators $\langle 0|q_i^2|0\rangle=\sum_\alpha (u^i_\alpha)^2 \langle 0|Q_\alpha^2|0\rangle+\sum_\alpha (v^i_\alpha)^2 \langle 0|P_\alpha^2|0\rangle$.
Assuming a regular behavior of the coefficient $(u^i_\alpha)^2$ and $(v^i_\alpha)^2$,
$\langle 0|q_i^2|0\rangle$ is typically the sum of terms
proportional to $\sim 1/\omega_\alpha$ when the mode frequency $\omega_\alpha$ is small.
}
which measures the strength of the deviations from the classical state. In such a case, the Néel ordered state is
stable with respect to quantum fluctuations, at least for large enough $S$.\footnote{This does not imply that the order should persists down to $S=\frac{1}{2}$.}
On the other hand, if the number of such modes grow like $N$, the 
average number of bosons diverge and the spin-wave expansion breaks down (the initial assumption that
$\langle 0|a^\dagger_i a_i|0\rangle$ is finite and small compared to $S$ cannot be satisfied).

At this point, a  route to obtain a QSL appears to look for a lattice where the classical model
has a sufficient number of ``soft'' modes, so that the zero point motion of the spins restore the rotation invariance and destroy the long range spin spin correlations. This condition is
realized on the kagome lattice, where indeed all numerical studies concluded to the absence of Néel
order in this system (at least for $S=\frac{1}{2}$). However, the semi classical spin wave theory described here breaks down.
As discussed in the next sections, QSL states in Mott insulators possess some internal topological properties
which are missed by the simple picture of a ``disordered'' state which would just be the
quantum analog of a high temperature phase.

\section{Lieb-Schultz-Mattis theorem, and Hastings's extension to $D>1$: ground state degeneracy in gapped spin liquids}
\label{sec:LSMH}

The Lieb-Schultz-Mattis theorem \cite{lsm61} was originally derived for spin chains and spin ladders \cite{affleck88,oya97}
and was recently extended to higher dimensions in an important work by Hastings \cite{hastings04}
(see also \cite{oshikawa00} for an intuitive topological argument valid in any dimension, and \cite{ns07}
for a mathematically rigorous proof).
It applies to spin Hamiltonians which are translation invariant in one direction (say $x$), 
have a conserved magnetization $S^z_{\rm tot}=\sum_i S^z_i$, and short range interactions. In addition, the model should have periodic boundary conditions in the $x$ direction.
Although more general interactions can easily be considered,\footnote{In particular, the interaction
can be anisotropic: $S^z_iS^z_j+\Delta (S^x_iS^x_j+S^y_iS^y_j)$, and an external magnetic field parallel to the $z$ direction can be present.} we concentrate for simplicity
on spin-$S$ Heisenberg models, written as in Eq.~\ref{eq:Heisenberg}
(with $J_{i,j}=J_{i+x,j+x}$ to respect the translation invariance).

\begin{figure}
\begin{center}
\includegraphics[height=3cm]{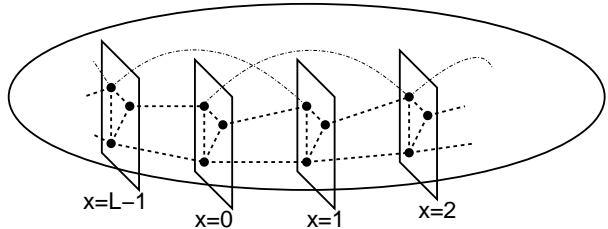}
\caption{A lattice model which is translation invariant and periodic in the $x$ direction can be viewed as a ring.
The interactions $J_{ij}$, indicated by dashed lines, are invariant in the $x$ direction but otherwise arbitrary.
In this example, each cross section has $C=3$ sites.
}\label{fig:ring}
\end{center}
\end{figure}

Following \cite{oshikawa00}, we define the {\it cross
section} as all the sites sitting at a given value of $x$.
By translation invariance, all cross sections are equivalent and contain $C$ sites (Fig.~\ref{fig:ring}).
In a spin chain, each cross section contains a single site. In an $n-$ leg spin ladder, $C=n$ sites. In a square lattice, $C=L_y$. On a $D$-dimensional lattice with $n$ sites per unit cell, $C=nL^{D-1}$, etc.
We note $L_x$ the system length in the $x$ direction, and therefore $CL_x$ is the total number of sites.
Finally we define $m^z=\frac{1}{CL_x}\langle0|S^z_{\rm tot}|0\rangle$ as the ground state magnetization
per site.

The theorem says that if $C(S+m^z)$ is {\it not an integer}, the ground state is either degenerate, or the spectrum has gapless excitations in the thermodynamic limit. In other words, if $C(S+m^z)\notin\mathbb{Z}$
the system {\it cannot have a unique ground state and a finite gap to excited states}
in the thermodynamic limit. Although the proof in 1D \cite{lsm61}
and  Oshikawa's topological argument \cite{oshikawa00} (Sec.~\ref{ssec:oshikawa}) are relatively simple,
the proof appears quite involved for $D>1$, and will not be discussed here.

What is the relation between the LSMH theorem and QSL ?
In most AF Heisenberg models on a finite-size lattice, $|0\rangle$ is a singlet and  $m^z=0$. If we focus on
the case $S=\frac{1}{2}$,  the theorem forbids a single ground state and a gap when $C$ is {\it odd}.
In  particular, if the lattice is two dimensional and describes a Mott insulator, the unit cell has an odd number $n$ of sites and  any odd $L_y$ can be chosen to get and odd $C=n L_y$ (note that the total number of sites is still even if $L_x$ is even).
If we assume that a {\it gapped} QSL is realized (for an example which fits in the LSMH conditions, see for instance \cite{bfg02}), its ground state must be {\it degenerate} (with periodic boundary conditions).
Usually, ground state degeneracies are the signature of some spontaneous symmetry breaking.
However, by definition, a QSL respect all lattice symmetries. The degeneracy imposed by the LSMH
theorem cannot be understood from this conventional point of view and is a hint that (gapped) QSL wave function possess some interesting topological properties, which correspond to the notion of ``topological order'' 
introduced by Wen \cite{wen89,wen91} for spin systems and Wen and Niu \cite{wn90} in the context of the fractional quantum Hall effect. As we will briefly discuss at the end, this topological degeneracy
is deeply related to the exotic nature of the elementary excitations in a QSL.\footnote{QSL have ``spinons'' excitations which carry a spin $\frac{1}{2}$ (like an electron) but no electric charge.}

\subsection{Oshikawa's topological argument}
\label{ssec:oshikawa}

Oshikawa's argument is somehow related to Laughlin's argument \cite{laughlin81}
for the quantization of the transverse conductivity in the quantum Hall effect.
First, a ``twisted'' version of the Hamiltonian is introduced:
\begin{equation}
 H_\theta=\frac{1}{2}\sum_{ij} J_{ij}\left[
	S_i^zS_j^z
	+\frac{1}{2}\left(
	e^{i\theta(x_i-x_j)/L_x} S_i^+S_j^- + {\rm H.c}
	\right)
\right]\label{eq:Htwisted}
\end{equation}
where $ 0 \leq x_i < L_x$ is the $x$-coordinate of site $i$.
It is simple to show that the spectra of $H_0$ and $H_{2\pi}$ are the same, since
the  unitary operator
\begin{equation}
 U=\prod_{i} \exp\left(2i\pi \frac{x_i}{L_x}S^z_i\right)
\end{equation}
maps $H_0$ onto $H_{2\pi}$:
\begin{equation}
	U H_0 U^{-1} = H_{2\pi}
	\label{eq:UH}
\end{equation}
(the calculation simply uses 
$e^{i\theta S^z_i}S^+_ie^{-i\theta S^z_i}=S^+_i e^{i\theta}$).

Starting with a spectrum of $H_0$ which is gapped, we further
assume that 
{\it the gap of $H_\theta$  remains  finite when $\theta$ goes from 0 to $2\pi$}.\footnote{Hastings
argument does not directly use $H_{\theta}$ for a finite $\theta$ and does not rely on this assumption.
This assumption is however reasonable by the fact that, under an appropriate choice of gauge (frame), $H=H_0$ and $H_\theta$ only differ for the
terms connecting the cross section at $x=L_x-1$ to the cross section at $x=0$ (boundary terms), and are identical in the bulk.
} On can follow
the ground state of $H_{\theta}$, which does not cross any other energy level as $\theta$ is varied.
Assuming that the ground state $|0\rangle$ of $H_0$ is unique and using the finite gap hypothesis, it must evolve
to the ground state of $H_{2\pi}$, denoted $|2\pi\rangle$.
Through Eq.~\ref{eq:UH}, both states are related: $|2\pi\rangle=U^{-1}|0\rangle$. 
However, the operator $U$ does not always commute with the translation operator $T$ and may change the momentum.
The precise relation is
\begin{equation}
	T U = U T \exp\left(2i\pi \frac{S^z_{\rm tot}}{L_x}\right)\exp\left(2i\pi C S\right) .
\end{equation}
The first phase factor,  also equal to $2\pi C m^z$, comes from the shift by $2\pi/L_x$ of the local rotation angles after a translation.
The second phase factor corrects the $2\pi$ jump of the rotation angle when passing from $x=L_x-1$ to $x=0$.
This relation implies that the momentum $k_0$ of $|0\rangle$ (defined by $T|0\rangle=e^{ik_0}|0\rangle$)
and the momentum $k_{2\pi}$ of $|2\pi\rangle=U^{-1}|0\rangle$ are related by
\begin{equation}
	k_{0}=k_{2\pi} +2\pi C(S+m^z) \label{eq:k}
\end{equation}
But $H_{\theta}$ is 
translation invariant (commutes with $T$)  and the momentum of each state (quantized for finite $L_x$) cannot change with $\theta$.
So $|0\rangle$ and $|2\pi\rangle$ have the same momentum and $k_0=k_{2\pi}\;[2\pi]$. From Eq.~\ref{eq:k}, we
get that $C(S+m^z)$ must be a integer.

\begin{figure}
\begin{center}
\includegraphics[height=3cm]{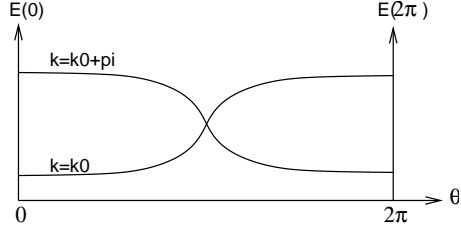}
\caption{Schematic spectrum of the twisted Hamiltonian (Eq.~\ref{eq:Htwisted}) as
a function the angle $\theta$, in the case where $C(S+m^z)$ is a half integer.
}\label{fig:twist}
\end{center}
\end{figure}

\section{Anderson's short range resonating valence-bond picture}
\label{sec:RVB}

\begin{figure}
\begin{center}
\includegraphics[height=3cm]{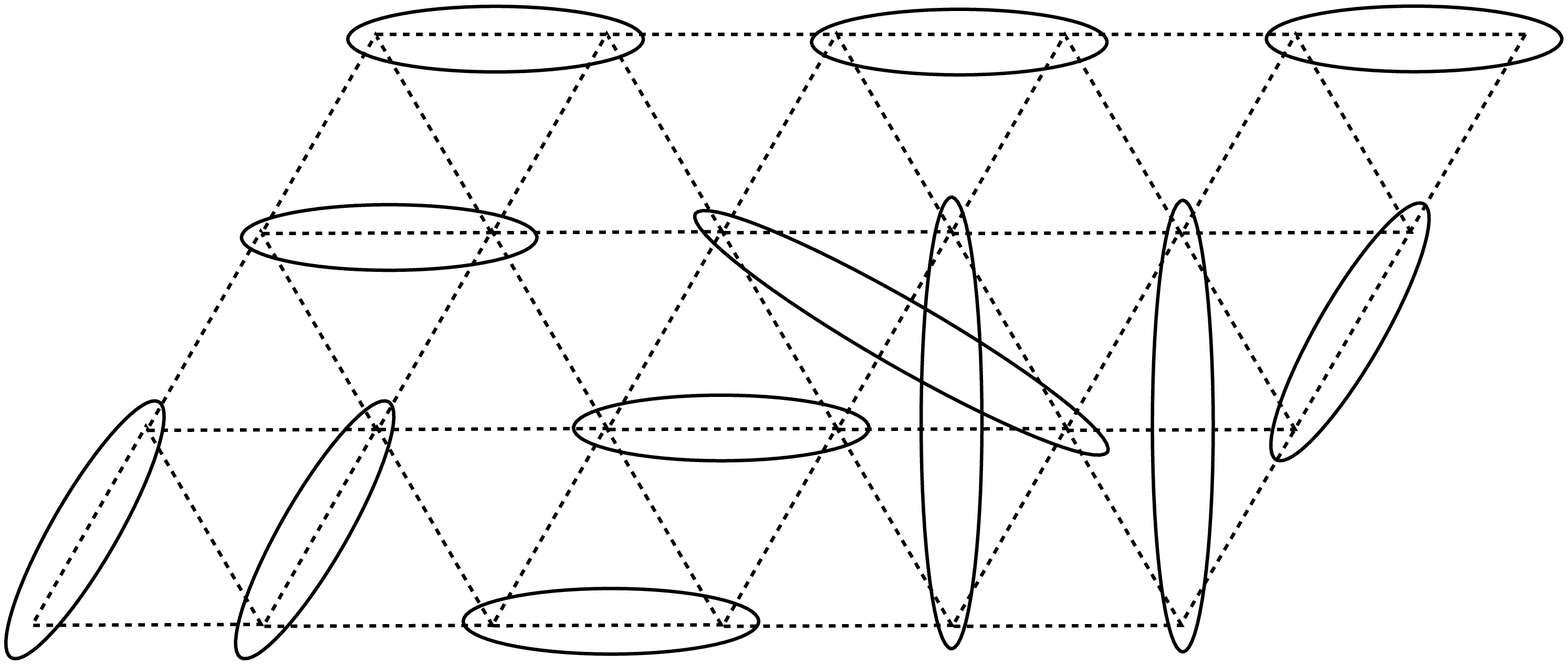}
\caption{A short range valence bond state on the triangular lattice. The singlet
pairs are marked with ellipses.}\label{fig:triVB}
\end{center}
\end{figure}

In the $1/S$ expansion, it is assumed that the spins experience small fluctuations about a well defined direction
and that spin-spin correlations are long ranged. This is of course incompatible with having a rotationally
invariant QSL state. To gain some intuition about what a QSL wave function may look like, it is instructive to start
from a completely opposite limit: a spin singlet state with extremely short range correlations. A {\it short range valence-bond} (VB) state is such a wave function, it is the direct product of $S=0$ states
$|[ij]\rangle=\frac{1}{\sqrt{2}}\left(|\uparrow_i\downarrow_j\rangle-|\downarrow_i\uparrow_j\rangle\right)$
on pairs of sites :
\begin{equation}
|{\rm VB}\rangle =|[i_0i_1]\rangle\otimes|[i_2i_3]\rangle\otimes|[i_4i_5]\rangle\otimes\cdots|[i_{N-1}i_N]\rangle
\end{equation}
where each site of the lattice appears exactly once (Fig.~\ref{fig:triVB}). Such a VB state is said to be short range if all
pairs of sites coupled in a singlet are at a distance $|{\bf r}_{i_{p}} - {\bf r}_{i_{p+1}}|$
smaller than or equal to some fixed  length $r_{\rm max}$(much smaller that the lattice size). The simplest case is $r_{\rm max}=1$,
where each spin forms a singlet with one of its nearest neighbors.

In a VB state, the spin-spin correlations are short ranged:
$\langle {\rm VB} | \vec S_i\cdot \vec S_j|{\rm VB}\rangle=0$ if $|{\bf r}_{i_{p}} - {\bf r}_{i_{p+1}}|>r_{\rm max}$. For a nearest neighbor Heisenberg model on a bipartite lattice, one can compare the (expectation value of the) energy of a nearest neighbor VB state, with that of the simple two-sublattice Néel state  $|\uparrow\downarrow\uparrow\downarrow\cdots\rangle$. The VB energy is $e_{\rm VB}=-J\frac{3}{8}$ per site
and the Néel one is $e_{\rm N}=-J\frac{z}{8}$, where $z$ is the number of nearest neighbors.
If the lattice is not bipartite but admits a three-sublattice classical ground states (with spins pointing at 120 degrees from each other), the energy of a classical Néel state is $e_{\rm N}=-J\frac{z}{16}$.
From this, we observe for instance that the VB energy is {\it lower}  than $e_{\rm N}$ on the kagome lattice.
More generally, this simple variational comparison shows that a low coordination $z$ and frustrated interactions (which increase the number of sublattices in the classical ground state) tend to favor VB states, and thus possible
QSL states.

In fact there are many (frustrated and Heisenberg-like) toy models where some/the nearest neighbor VB states are {\it exact} ground states. The most famous example
is the Majumdar-Gosh model \cite{mg69}.
Consider the spin-$\frac{1}{2}$ Heisenberg chain with first- ($J_1$) and second- ($J_2$)
neighbor couplings. At $J_1=2J_2>0$ we have
\begin{equation}
 H_{\rm MG}=2\sum_i \vec S_i \cdot \vec S_{i+1}+\sum_i \vec S_i \cdot \vec S_{i+2}
\end{equation}
and the (two-fold degenerate) ground states are exactly given:
\begin{eqnarray}
|a\rangle &=&\cdots\otimes|[01]\rangle\otimes|[23]\rangle\otimes|[46]\rangle\otimes\cdots \\
|b\rangle &=&\cdots\otimes|[12]\rangle\otimes|[34]\rangle\otimes|[56]\rangle\otimes\cdots
\end{eqnarray}
The proof can be done three steps. First, the Heisenberg Hamiltonian on
three site $H_{ijk}=\vec S_i \cdot \vec S_j + \vec S_j \cdot \vec S_k + \vec S_k \cdot \vec S_i$
is written as $H_{ijk}=\frac{1}{2}(\vec S_i+\vec S_j+\vec S_k)^2-\frac{9}{8}$.
In this form, proportional to the square of the total spin, it is clear that 
the eigenvalues of $H_{ijk}$ are $\frac{1}{2}S(S+1)-\frac{9}{8}$ with $S=\frac{1}{2}$ or $S=\frac{3}{2}$
(the only possible values of $S$ for three spin-$\frac{1}{2}$).
So, if the sites $ijk$ are in a $S=\frac{1}{2}$ state, they minimize exactly $H_{ijk}$.
Second, one expresses the Majumdar-Gosh Hamiltonian
as
\begin{equation}
 H_{\rm MG}=\sum_i H_{i-1,i,i+1}.\label{eq:HMG}
\end{equation}
Finally, one remarks that the dimerized states $|a\rangle$ and $|b\rangle$
always have one singlet among the sites $i-1,i,i+1$, which are therefore in a $S=1/2$ state.
We conclude that $|a\rangle$ and $|b\rangle$ minimize all the terms in Eq.~\ref{eq:HMG}
and are thus ground states of $H_{\rm MG}$.

The Majumdar-Gosh model is the simplest model of a family of spin models
where exact VB ground states can be found.\footnote{
There  exists a general method for constructing an $SU(2)$ symmetric spin model with short range interactions
such that all the nearest neighbor VB states are ground states \cite{klein82}.
Building on this idea, it was possible to construct $SU(2)$ symmetric spin-$\frac{1}{2}$ models
(with short ranged interaction) with a gapped QSL ground state \cite{rms05}. Although complicated,
these models are among the very few examples where the ground state is well
established to be a short ranged resonating VB liquid.
} For instance,
the Husimi cactus \cite{cd94} is a lattice constructed as a tree (no loops) of corner sharing triangles.
Its geometry is locally similar to the kagome lattice but it has no closed loop (except of course for the triangles themselves). The argument above (writing the Hamiltonian as a sum of $H_{ijk}$) directly generalizes to this case and shows that any nearest neighbor VB state is a ground state.
One can also mention the 2D Shastry-Sutherland Heisenberg model \cite{ss81}, where a particular nearest neighbor VB is the unique ground state, and which has an experimental realization in SrCu$_2$(BO$_3$)$_2$ \cite{k99}.

So far, we do not yet have any gapped {\it liquid} state.\footnote{The ground states
of $H_{MG}$ spontaneously break the translation symmetry. On the Husimi cactus, the ground state is highly degenerate. The Shastry-Sutherland ground state does not break any symmetry (the ground state is unique), but the lattice has an {\it even} number of spins per unit cell and
should be considered as a band insulator in our classification.} To obtain a qualitative idea of how VB states can
be the building blocks of a gapped QSL, we will briefly explain the short range resonating valence bond (RVB)
picture proposed by Anderson \cite{anderson73}. If we exclude the toy models discussed above, a VB state is
generally not an eigenstate of the Heisenberg  model. Starting from a nearest neighbor VB state, the
Heisenberg Hamiltonian will induce some dynamics among the VB states. If we take the kagome example, a nearest neighbor VB state inevitably contains some ``defect'' triangles {\it without} any singlet.\footnote{Whatever
the nearest neighbor VB state, exactly 1/4 of the triangles have no singlet bonds.}
While the term $H_{ijk}$ leaves the VB state unchanged if the corresponding triangle has a singlet bond, the
three
VB touching $i$, $j$ and $k$ will be moved by  $H_{ijk}$ if $(ijk$) is a defect triangle. The ground state can be viewed as a linear combination of (many) VB configurations (not necessarily nearest neighbor).
Anderson suggested that, with appropriate interactions and lattice geometry,
the ground state wave function could be ``delocalized'' over a large part of the subspace spanned by short range VB states.
By forming a linear superposition of a large number of very different VB states, the system may restore
all the lattice symmetries (which are broken by an individual VB state) and form a QSL.

A more formal  approach to this idea will be discussed 
in Sec.~\ref{sec:SB}, but this picture can already be used to anticipate the nature of the magnetic excitations
in such a short range RVB liquid. To this end, we first consider a 2D model where one ground state
is equal to (or dominated by) {\it one} particular VB state.
Contrary to the Anderson's RVB liquid, the wave function is localized in the vicinity of one particular VB state.
It can be thought as a 2D analog of the Majumdar Gosh chain,
where the ground state is a spatially regular arrangement of singlet bonds.
Many 2D models are known to realize such VB {\it crystals} (VBC) \cite{ml05}, and we refer to Ref. \cite{glkm07} for a recent example
where the exact ground states are known.
In a VBC, a finite energy excitation can be created by replacing a singlet bond by a triplet ($S=1$), with an energy cost proportional to $J$.
But is it possible to construct two {\it separated} spin-$\frac{1}{2}$ excitations in such a system ?
As a trial state, one can place two remote spins ``up'' (two spinons excitations) at sites 0 and $i$.
Then, to minimize the energy, the regular VB structure of
the ground state should be reconstructed as much as possible. However, due to the   spinons,
the regular pattern cannot be fully reconstructed between 0 and $i$, and a "string" of misaligned VB is unavoidable.
The unpaired spins behave as a topological defect in the crystalline order.
So, two remote spinons perturb the ordered VB background, not only in their vicinity, but all the way between them. They lead to an energy cost which is proportional to their separation.\footnote{The situation is very different in 1D. In the Majumdar-Gosh model, one can get a {\it finite energy} state
with two remote spinons by introducing a domain wall in the dimerization pattern in 0 and $i$.}
So, isolated spinons are not finite energy excitations in a VBC. The ordered VB background is a medium which confines the spinons in pairs. Since an RVB state should instead be viewed as a {\it liquid} (no broken symmetry, no long range order), it is reasonable to expect the spinons
to be able to propagate as  independent particles.
As we will see in the next section, the proper way to address this question of confinement and deconfinement
of spinons is to understand the emergence of gauge degrees of freedom in these systems.

\section{Schwinger bosons, large-$\mathcal N$ limit, and $\mathbb{Z}_2$ topological phase}
\label{sec:SB}

\subsection{Schwinger bosons representation}
The spin wave approach is a large-$S$ approach and is unable to capture highly quantum states which are rotationally symmetric, such as RVB wave functions.
From the discussion of Sec.~\ref{sec:RVB}, it is natural to look for a description 
in terms of {\it singlet fields} leaving on bonds, and able to describe the presence or absence of a singlet
between two sites. Such variables appear naturally when using the Schwinger boson
representation of the spin operators \cite{aa88,auerbachbook}.

At each site,
two types of bosons carrying a spin ``up'' and ``down'' are introduced: $a^\dagger_{i\uparrow}$ and
$a^\dagger_{i\downarrow}$, and the spin operators are represented as bilinears in the boson creation and
annihilation operators
\begin{eqnarray}
S^z_i=\frac{1}{2}\left(
a^\dagger_{i\uparrow}a_{i\uparrow}
-a^\dagger_{i\downarrow}a_{i\downarrow}
\right)
\;\;,\;\;
S^+_i=a^\dagger_{i\uparrow}a_{i\downarrow}
\;\;,\;\;
S^-_i=a^\dagger_{i\downarrow}a_{i\uparrow}
\end{eqnarray}
With these relations, the commutation relations
$[S^\alpha_i,S^\beta_i]=i\epsilon^{\alpha\beta\delta}S^\delta_i$
are automatically verified. The total spin reads
$
\vec S_i^2=\frac{n_i}{2}\left(\frac{n_i}{2}+1\right)
$,
where $n_i=a^\dagger_{i\uparrow}a_{i\uparrow}
+a^\dagger_{i\downarrow}a_{i\downarrow}$ is the total number of bosons at site $i$.
To fix the length of the spins, the following constraint must therefore be imposed on physical states:
\begin{equation}
a^\dagger_{i\uparrow}a_{i\uparrow}
+a^\dagger_{i\downarrow}a_{i\downarrow}=2S
\label{eq:cons}
\end{equation}

With this representation,\footnote{Fermions can also be used,
leading to other very interesting
theories for (gapped of gapless) QSL \cite{a85,am88,gaplessQSL}.} the Heisenberg interaction is of degree four in the boson operators and can be written
\begin{eqnarray}
\vec S_i \cdot \vec S_j = S^2-\frac{1}{2} (A_{ij})^\dagger A_{ij} \label{eq:SSA2}\\
{\rm with }\;\;A_{ij}=a_{i\uparrow}a_{j\downarrow}-a_{i\downarrow}a_{j\uparrow}.
\end{eqnarray}
The bond operators $A_{ij}^\dagger$ behave as
a singlet creation operators: $A_{ij}^\dagger$, when applied onto the boson vacuum, creates
a spin singlet  $|\uparrow_i\downarrow_j\rangle-|\downarrow_i\uparrow_j\rangle$ and,
from Eq.~\ref{eq:SSA2}, $A_{ij}^\dagger A_{ij}$ is proportional to the number (0 or 1) of a singlet
between sites $i$ and $j$. In addition, $A_{ij}$ is invariant under rotations:
redefining the bosons by an $SU(2)$ matrix $P$:
$\left[\begin{array}{c}a_\uparrow\\a_\downarrow\end{array} \right]
\to P \left[\begin{array}{c}a_\uparrow\\a_\downarrow\end{array} \right]$
leaves $A_{ij}$ unchanged.\footnote{
$A_{ij}$ can be written using the $2\times 2$ antisymmetric tensor $\epsilon=\left[\begin{array}{cc}0&-1\\1&0\end{array} \right]$:
$A_{ij}=\sum_{\sigma,\sigma'=\uparrow,\downarrow} \epsilon_{\sigma\sigma'} a_{i\sigma} a_{j\sigma'}$. 
The rotation invariance of $A_{ij}$ follows from the fact that any $P\in SU(2)$ satisfies
$P^t \epsilon P=\epsilon$.
}

\subsection{Mean field approximation}

Arovas and Auerbach \cite{aa88}
suggested
an approximation in which the interaction is decoupled using mean-field expectation values
\begin{equation}
 A_{ij}^\dagger A_{ij} \longrightarrow A_{ij}^\dagger \langle A_{ij} \rangle
				+ \langle A_{ij}^\dagger\rangle  A_{ij}
				- |\langle A_{ij}^\dagger\rangle|^2 
	\label{eq:MF}
\end{equation}
and to replace the constraint (Eq.~\ref{eq:cons}) by a condition on the {\it average} number of boson per site
\begin{equation}
	\langle a^\dagger_{i\uparrow}a_{i\uparrow}
+a^\dagger_{i\downarrow}a_{i\downarrow} \rangle =2S. \label{eq:n2S}
\end{equation}
By this replacement, the Hamiltonian becomes quadratic in the boson operator
\begin{eqnarray}
	H	\longrightarrow H_{\rm MF}[Q_{ij}^0,\lambda_j^0]=-\frac{1}{2}\sum_{ij} 
		\left(
			A_{ij}^\dagger Q_{ij}^0+ \bar{ Q_{ij}^0}  A_{ij}
		\right)\nonumber\\
		-\sum_i \lambda_i^0 \left(
			a^\dagger_{i\uparrow}a_{i\uparrow}
			+a^\dagger_{i\downarrow}a_{i\downarrow} -2S
		\right)
	+{\rm cst}.
\end{eqnarray}
A chemical potential $\lambda_i^0$ has been introduced at each site to tune the boson densities
so that they satisfy Eq.~\ref{eq:n2S}.
The mean field Hamiltonian $H_{\rm MF}$ (and thus its ground state $|0\rangle$) depends
on the complex parameters $Q_{ij}^0$ (one for each pair of sites $ij$ where $J_{ij}\ne 0$). These parameters have to 
be adjusted to satisfy the self-consistency conditions on each bond
\begin{equation}
	 Q_{ij}^0 = \frac{1}{2}J_{ij}
	\langle 0 | a_{i\uparrow}a_{j\downarrow}-a_{i\downarrow}a_{j\uparrow}|0\rangle.
	\label{eq:SC}
\end{equation}
As in the spin wave approach, the Heisenberg model has been reduced to a
quadratic boson model (here with some self consistency conditions). However,
the crucial difference is that the present formalism does not impose any preferred spin direction:
giving a finite expectation value $A_{ij}^0\ne 0$ to the operator $A_{ij}$ does not
break the $SU(2)$ symmetry -- which is a necessary condition to describe a QSL.

Generally speaking, two family of solutions can be found at this mean field level.
In the first class, favored when $S$ is large, the Schwinger boson {\it Bose-condense} in some particular mode. Because they carry a spin index, such condensate state (spontaneously) breaks the $SU(2)$ symmetry. These solutions describe Néel states with long range spin-spin correlations. In such cases, the Schwinger boson mean-field theory is essentially equivalent  to the spin wave approach (Sec.~\ref{sec:SW}).

The second class corresponds to (mean field) QSL states.
There, the ground state is rotationally invariant, and the Bogoliubov quasi particles obtained by diagonalizing $H_{\rm MF}$ are gapped.
Since the
corresponding creation operators, $b_{\uparrow,\alpha}$ and $b_{\downarrow,\alpha}$, are linear combinations
of the original bosons, these excitations also carry a spin $\frac{1}{2}$.
The most important question is  whether the existence
of  these deconfined  (free  in the mean field approximation) spinons    is an artifact  of the
mean field  approximation, or if they could survive in some Heisenberg spin model.  In    the first a  case,  the  inclusion   of
the fluctuations that were neglected would  confine  the
spinons and would deeply   change the  nature of the ground state.
The mean-field picture of a fully symmetric state with non interacting spinons excitation is then qualitatively incorrect.
Another possibility is that the spinons remains deconfined, even in
presence of fluctuations. In that case, the
mean-field  approximation  is   a very useful  starting  point.
We will discuss in Sec.~\ref{ssec:Z2} a scenario where it is the case.
But before, we need to introduce the basic formalism that is needed to
describe the fluctuations about the mean field solution, and
emergence of gauge degrees of freedom in the system.
The central question concerning the long distance and low energy properties of the system will be whether these gauge degrees of freedom confine or not the spinons.

\subsection{Large $\mathcal N$, saddle point}

To discuss the role of the fluctuations neglected in Eq.~\ref{eq:MF},
it is necessary to formulate the mean field approximation
as a saddle point approximation in path integral formulation of the model.
It will then be possible to identify the
structure of the most important fluctuations about the saddle point.
To do so, one duplicates $\mathcal N$ times the two species  of bosons ($\uparrow$ and $\downarrow$). In addition to the site and up/down indices $\sigma$, the boson operators
now carry an additional ``flavor'' index $m=1,\cdots,\mathcal{N}$.
The Hamiltonian and the constraint are then generalized to
\begin{eqnarray}
	H&=&-\frac{1}{2\mathcal{N}}\sum_{ij}J_{ij} A_{ij}^\dagger A_{ij} \\
	A_{ij}&=&\sum_{m=1}^\mathcal{N} a_{im\uparrow}a_{jm\downarrow}-a_{im\downarrow}a_{jm\uparrow}
\end{eqnarray}
and
\begin{equation}
\sum_{m=1}^\mathcal{N}
a^\dagger_{im\uparrow}a_{im\uparrow}
+a^\dagger_{im\downarrow}a_{im\downarrow}=2\mathcal{N}S. \label{eq:Ncons}
\end{equation}
For $\mathcal{N}=1$, this model is Heisenberg model with $SU(2)$ symmetry.
For $\mathcal{N}>1$, this model has an enlarged symmetry given by the group $Sp(\mathcal{N})$.\footnote{
The simplectic group of $2\mathcal{N}\times 2\mathcal{N}$ matrices $Sp(\mathcal{N})$ is the set of matrices $P$ which satisfies
$P^t \mathcal{J} P = \mathcal{J}$, where
$\mathcal{J}=\left[
\begin{array}{ccccc}
0	&1	&	&	&	\\
-1	&0	&	&	&	\\
	&	&\ddots	&	&	\\
	&	&	& 0	& 1	\\
 	&	&	& -1	& 0	
\end{array}\right]$
generalizes the antisymmetric $\epsilon$ tensor.
}
$S$ is a parameter of the model, and is no longer related to
a representation of $SU(2)$ if $\mathcal{N}>1$.
The bond operator $A_{ij}$ is a sum over all the  flavors.
For this reason, in the limit where $\mathcal{N}$ is very large, the fluctuations
of $A_{ij}$ become negligible compare to its expectation value and the approximation
made in Eq.~\ref{eq:MF} becomes exact.

A formal way to establish this result is to adopt a formulation of model where the
partition function $\mathcal{Z}={\rm Tr}\left[e^{-\beta H}\right]$ at temperature $T=\beta^{-1}$ is expressed
as a coherent state path integral over complex variables $z_{im\sigma}(\tau)$ (in correspondence with the
boson operators $a_{im\sigma}$) which are periodic functions of the imaginary
time $\tau\in[0,\beta [$. In this formalism the partition function reads
\footnote{
For an introduction to the path integral formalism in this context of
quantum magnetism, see for instance Ref.~\cite{auerbachbook}.
We sketch the main steps of the derivation in the case of a single bosonic mode $[a,a^\dagger]=1$.
For any complex number $z$, a coherent state $|z\rangle=e^{z a^\dagger}|0\rangle$ is defined.
These states satisfy: $a|z\rangle=z|z\rangle$, $\langle z |z'\rangle=e^{\bar z z'}$
and the resolution of the identity $\frac{1}{\pi}\int d^2 z\;|z\rangle\langle z|e^{-|z|^2}={\bf 1}$.
On writes the partition function as a product over $N_\tau$ imaginary time steps
$\mathcal{Z}={\rm Tr}\left[
	e^{-d\tau H}e^{-d\tau H}\cdots
	\right]=\lim_{N_{\tau}\to\infty}{\rm Tr}\left[
	(1-d\tau H)(1-d\tau H)\cdots
	\right]$
with $d\tau=\beta/N_{\tau}$. Then, the identity is inserted at each step:
$\mathcal{Z}=\lim_{N_\tau\to\infty}\int\left(\prod_{\tau=1}^{N_{\tau}} d^2 z_\tau\right)
e^{-|z_1|^2}\langle z_1 | 1-d\tau H|z_{N_{\tau}} \rangle e^{-|z_{N_\tau}|^2} \langle z_{N_\tau} |1-d\tau H |z_{N_\tau-1} \rangle
\cdots
e^{-|z_2|^2}\langle z_{2} | 1-d\tau H|z_{1} \rangle 
$. Next, we write
$e^{-|z_i|^2} \langle z_i |1-d\tau H |z_{i-1} \rangle
\simeq \exp{\left[
-\bar z_i(z_i-z_{i-1})-d\tau H(\bar z_i,z_{i-1})
\right]}$, where the complex number $H(\bar z,z')=\langle z'|H |z \rangle $
is obtained by writing the Hamiltonian in a normal-ordered form an replacing $a^\dagger$
by $\bar z$ and $a$ by $z'$.
Taking the continuous time limit $d\tau\to 0$ is formally written as
$z_i-z_{i-1}\to\partial_\tau z(\tau)d\tau$
and finally leads to $\mathcal{Z}=\int \mathcal{D}[z] \exp(-\int_0^\beta Ld\tau)$ with
the Lagrangian
$L=\bar z(\tau) \partial_\tau z(\tau) + H(\bar z(\tau),z(\tau))$.
}
\begin{eqnarray}
 \mathcal{Z}&=&\int \mathcal{D}[z_{im\sigma}(\tau),\lambda_i(\tau)]
	\exp\left(-\int_0^\beta L_0\;d\tau\right) \\
 L_0&=&\sum_{i\;m\;\sigma} \bar z_{im\sigma} \partial_\tau z_{im\sigma}
	-\frac{1}{2\mathcal{N}}\sum_{ij}J_{ij} A_{ij}^\dagger A_{ij} \nonumber \\
	&&+i\sum_{i\;m} \lambda_i\left(
		\bar z_{im\uparrow}z_{im\uparrow}+\bar z_{im\downarrow}z_{im\downarrow}-2S
	\right)
\\
	A_{ij}&=&\sum_{m=1}^\mathcal{N} \left(
		z_{im\uparrow}z_{jm\downarrow}-z_{im\downarrow}z_{jm\uparrow}
	\right),
\end{eqnarray}
where a Lagrange multiplier $\lambda$ has been introduced at each lattice site and each time step
to enforce the constraint (Eq.~\ref{eq:Ncons}) exactly (to simplify the notations,
the $\tau$ dependence of all
fields is implicit).

Now, a Hubbard-Stratonovich transformation is performed :
\begin{eqnarray}
 \mathcal{Z}&=&\int \mathcal{D}[z_{im\sigma}(\tau),\lambda_i(\tau),Q_{ij}(\tau)]
	\exp\left(-\int_0^\beta L_1\;d\tau\right) \\
 L_1&=&\sum_{i\;m\;\sigma} \bar z_{im\sigma} \partial_\tau z_{im\sigma}
	+\sum_{ij}\left(
	\frac{2\mathcal{N}}{J_{ij}} |Q_{ij}|^2 - \bar Q_{ij} A_{ij} - Q_{ij} \bar A_{ij}
	\right) \nonumber \\
	&&+i\sum_{i\;m} \lambda_i\left(
		\bar z_{im\uparrow}z_{im\uparrow}+\bar z_{im\downarrow}z_{im\downarrow}-2S
	\right)
	\label{eq:L1}
\end{eqnarray}
This new formulation involves an additional complex field $Q_{ij}$ on each bond.
The equivalence of $L_1$ with the initial Lagrangian $L_0$ can simply be checked by performing
the Gaussian integrations over $Q_{ij}(\tau)$ for each bond and each time step:
$\int \mathcal{D}[Q_{ij}(\tau)] \exp\left(-\int_0^\beta L_1\;d\tau\right)
= \exp\left(-\int_0^\beta L_0\;d\tau\right)$ (up to a multiplicative constant).
At this point, the $\mathcal N$ flavors of particles are no longer coupled to each other, but are
coupled to a common bond field $Q_{ij}$. So, for a fixed space-time configuration of $Q$, we
have $\mathcal{N}$ independent copies of the same boson system.
In addition, the Lagrangian $L_1$ is now quadratic in the $z$ variable.
We note $G_{Q,\lambda}^{-1}$ the corresponding quadratic form, a big
matrix which has space ($i$), time ($\tau$), spin ($\sigma$) and complex conjugacy ($z$ versus $\bar z$) indices (but no flavor index), and depends on the auxiliary field $Q$ and $\lambda$.
$L_1$ is then
\begin{eqnarray}
  L_1&=&\sum_{ij}
	\frac{2\mathcal{N}}{J_{ij}} |Q_{ij}|^2 -2i\mathcal{N}S\sum_{i} \lambda_i \nonumber \\
	&&+\sum_m 
	\left[\bar z_{i\sigma}(\tau) ; z_{i\sigma}(\tau)\right]
	G_{Q,\lambda}^{-1}
	\left[\begin{array}{c}z_{j\sigma'}(\tau')\\\bar z_{j\sigma'}(\tau')\end{array}\right] 
\end{eqnarray}
Performing the Gaussian integral over the $z$ fields is now  simple,
as it gives $({\rm det}[G])^\mathcal{N}$, also equivalent
to
$e^{\mathcal{N}{\rm Tr}[\log(G)]}$.
The partition function is now expressed
as a path integral with the fields $Q$ and $\lambda$ only, but with a complicated
non-Gaussian weight:
\begin{eqnarray}
\mathcal{Z}&=&\int \mathcal{D}[z_{i\sigma}(\tau),\lambda_i(\tau),Q_{ij}(\tau)]
	\exp\left(-\mathcal{N} \int_0^\beta L_2\;d\tau\right) \\
 L_2&=&	+\sum_{ij}
	\frac{2}{J_{ij}} |Q_{ij}|^2 -2iS\sum_{i} \lambda_i+{\rm Tr}[\log(G_{Q,\lambda})]
\end{eqnarray}
Here, the flavor indices $m$ have disappeared and $\mathcal{N}$ only appears a global multiplicative factor in the action.
With this formulation of the $Sp(\mathcal N)$ ``spin'' model, it is  clear that, in the limit $\mathcal{N}\to \infty$
the partition function will be dominated by the
configurations $(Q^0,\lambda^0)$
which are {\it saddle points} of the action
$\mathcal{S}[Q,\lambda]=\int_0^\beta L_2\;d\tau$.
In other words {\it the fluctuations of $Q_{ij}$ and $\lambda_i$ are frozen when $\mathcal{N}\to\infty$}.
Such saddle points are obtained by requiring
\begin{eqnarray}
 	\left.\frac{\partial \mathcal{S}}{\partial \lambda_i(\tau)}\right|_{Q^0,\lambda^0}=0
	\;\;,\;\;
	\left.\frac{\partial \mathcal{S}}{\partial Q_{ij}(\tau)}\right|_{Q^0,\lambda^0}=0. \\
\end{eqnarray}
and in most cases they are found to be time independent $Q_{ij}^0(\tau),\lambda_i^0(\tau)\to Q_{ij}^0,\lambda_i^0$.
The equations above can then be shown to be equivalent to the self consistency conditions of Eqs.~\ref{eq:n2S} and \ref{eq:SC}, with $Q_{ij}^0=\frac{J_{ij}}{2\mathcal{N}}\sum_m
\langle0|
a_{im\uparrow}a_{jm\downarrow}-a_{im\downarrow}a_{jm\uparrow}
|0\rangle
$.

\subsection{Fluctuations about a saddle point and gauge invariance}

We are now ready to discuss the fluctuations that are present when $\mathcal{N}$ is finite,
where the field $Q_{ij}(\tau)$ is  able to fluctuate around its mean field
value $Q_{ij}^0$. Treating all the possible fluctuations is certainly very difficult, as
it would amount to solve the original spin problem.
A possible approach is to compute perturbatively the first $1/\mathcal{N}$ corrections
to the mean field results \cite{aa88}.
However, this can miss some important effect (instabilities) which are not perturbative in $1/\mathcal{N}$,
and will generally not shed light on the issue of spinon confinement that we are interested in.
Instead, as in \cite{rs89,rs91}, we will examine the qualitative structure of the fluctuation modes which are important for the long distance properties of the system.
In particular, we would like to know if some fluctuations could confine the spinons
(in which case the mean field picture is incorrect), or if the QSL state is stable at finite $\mathcal{N}$.
As we will see, there are some fluctuations modes which are described by a gauge field \cite{rs89,rs91}
and mediate some (possibly long ranged) interaction between the spinon. The dynamics of this gauge field is therefore crucial to the physics of the spin system.
In some cases this gauge field will be in a confining phase, and the $\mathcal{N}=\infty$ limit (where the fluctuations
are frozen out) does not represent the physics of the finite $\mathcal{N}$ models \cite{rs89}. In some other situations,
the gauge field has a deconfined phase and a QSL state with elementary spinon excitation is possible \cite{rs91}.

First, it should be noticed that the description of the spin operators with Schwinger bosons
is redundant in the sense that an arbitrary local change of phase in the boson operators does not change the 
physical spin operators. In the path integral formulation, this becomes
a full space-time gauge invariance. The Lagrangian $L_1$ (Eq.~\ref{eq:L1})
is invariant under
\begin{eqnarray}
	z_{im\sigma}(\tau)&\longrightarrow&	e^{i\Lambda_i(\tau)} z_{im\sigma}(\tau) \label{eq:zL}\\
	Q_{ij}(\tau)&\longrightarrow&		e^{i(\Lambda_i(\tau)+\Lambda_j(\tau))} Q_{ij}(\tau) \label{eq:QL}\\
	\lambda_i(\tau)&\longrightarrow&	\lambda_i(\tau)-\partial_\tau \Lambda_i(\tau)
\end{eqnarray}
where $\Lambda_i(\tau)$ is some  arbitrary angle at each site and time step.

However, this local $U(1)$ gauge invariance is broken to a smaller
invariance group in the vicinity of a saddle point $(Q^0,\lambda^0)$.
This can be illustrated the simpler context of a classical ferromagnetic  Heisenberg model.
A ground state  is magnetized in one particular direction and thus breaks the $O(3)$ symmetry of the Hamiltonian.
The theory for the (transverse) spin  {\it deviations} around this ferromagnetic state has an $O(2)$
symmetry, an  not $O(3$).
The situation is similar for the fluctuations of the bond field $Q_{ij}$.
Although the model has a local $U(1)$ gauge invariance, {\it the action describing the fluctuations around $Q^0_{ij}$
have a lower invariance group}. In the ferromagnet example, we look at the rotations under which the ground state is unchanged. Similarly, we look for the gauge transformations which leave $Q^0_{ij}$ unchanged.
These transformations form the {\it invariant gauge group} (IGG) of the saddle point, a concept introduced by X.~G. Wen \cite{wen02}.
A gauge transformation $i\mapsto\Lambda_i$ belongs to the IGG of $Q^0_{ij}$
if it is static and satisfies
\begin{equation}
 	Q_{ij}^0 = Q_{ij}^0 e^{i(\Lambda_i+\Lambda_j)}
	\label{eq:IGG}
\end{equation}
If the lattice made of the bonds where $Q^0_{ij}$ is non zero is {\it bipartite},
it is easy to show that $\Lambda_i=\theta$ on sublattice A and $\Lambda_i=-\theta$
on sublattice $B$ satisfies Eq.~\ref{eq:IGG} for any (global) angle $\theta$.
In such a case, the IGG is isomorphic to $U(1)$. On the other hand,
if the lattice of the bonds where $Q^0_{ij}\ne0$
is {\it not} bipartite, the IGG is isomorphic to $\mathbb{Z}_2$, since
$\Lambda_i=\pi$ and $\Lambda_i=0$ are the only two solutions to Eq.~\ref{eq:IGG} when $Q_{ij}^0\ne0$.

The general result \cite{wen02} is that, among the fluctuations
around the saddle point $Q^0$, some modes
are described by a {\it gauge field}. with a gauge group given by the IGG.
We will illustrate this result in the simple case  IGG$=\mathbb{Z}_2$.\footnote{The cases where IGG$=U(1)$
are generically unstable saddle points: the gauge fluctuations lead to spinon confinement, and lattice symmetry
breaking (VBC) when $S=\frac{1}{2}$ \cite{rs89}. This will not be discussed here.}

\subsection{$\mathbb{Z}_2$ gauge field}
\label{ssec:Z2}
If the IGG is $\mathbb Z_2$, the important fluctuations turn out to be 
fluctuations of the {\it sign} of $Q_{ij}$. We therefore parametrize these
fluctuations  in the following
way
\begin{equation}
 	Q_{ij}(\tau) = Q_{ij}^0 \;e^{i\mathcal{A}_{ij}(\tau)} 
	\;\;,\;\;
	\mathcal{A}_{ij}(\tau)\in \{0,\pi\}. \label{eq:A}
\end{equation}
where the field $\mathcal{A}_{ij}$ will play the role of a ``discrete'' ($\mathbb Z_2$) vector
potential living on the links of the lattice (pairs of sites where $Q_{ij}^0\ne0)$.

Doing the integration over all the other fluctuation modes
(amplitude fluctuations the bond field $Q_{ij}$, fluctuations of $\lambda_i$, etc.)
in order to  obtain an effective action for $\mathcal{A}_{ij}$ and the bosons
$z_{i\sigma}$ only\footnote{From now on, we go back to $\mathcal{N}=1$ a drop the flavor index $m$ for simplicity.}
is formally possible, but it is of course a very difficult task in practice. One can instead determine the symmetry constraints, and, in a Landau-Ginzburg type of approach, construct the simplest action compatible with these symmetries.

For this, we consider the (static) local gauge transformation
$i\mapsto\Lambda_i$ with the restriction $\Lambda_i\in \{0,\pi\}$. Because
$\mathcal{A}_{ij}$ is defined modulo $2\pi$, $-\Lambda_j$ is equivalent to
$+\Lambda_j$ and the transformation rules take the usual form (except for the discrete nature
of $\mathcal A_{ij}$):
\begin{eqnarray}
 	z_{i\sigma}&\longrightarrow& e^{i \Lambda_i} z_{i\sigma} \\
	\mathcal{A}_{ij}&\longrightarrow & \mathcal{A}_{ij} + \Lambda_i-\Lambda_j.
\end{eqnarray}
These local transformations form a very large symmetry group (2 to the power of the number of lattice sites)
and severely constrain the effective Hamiltonian for these degrees of freedom.
Because of this invariance,
a term like $\mathcal{A}_{ij}$, $\mathcal{A}_{ij}^2$ or even $\cos(\mathcal{A}_{ij})$ cannot appear as an energy term.\footnote{In the same way, a a mass term like the square of
the vector potential $A_{\mu\nu}^2$
is forbidden by gauge invariance in conventional electromagnetism.}
Instead, only the products of $e^{i\mathcal{A}_{ij}}$ on closed loops are gauge invariant. As a circulation of the a vector potential, these loop terms are the analog of the magnetic flux in  electromagnetism.
Such products can thus appear in an effective description of the fluctuations about the
mean field solution.
Terms like $\mathcal{E}_{ij}=\partial_\tau \mathcal{A}_{ij}+\lambda_i-\lambda_j$, which are equivalent
to the electric field, are also gauge invariant. 
As for the couplings to the bosons, the coupling to $\mathcal{A}$ allowed by 
the gauge invariance (an spin-rotations) are of the type
$\bar z_{i\sigma}\;e^{i\mathcal{A}_{ij}}\;z_{j\sigma}$.

\subsection{A simple effective model}
We can combine the gauge invariant terms above into a simple Hamiltonian
which can phenomenologically, when IGG=$\mathbb Z_2$, describe the gauge fluctuations about a
saddle point and their effect on the spinons:
\begin{eqnarray}
 H&=&-K\sum_{\Box} \sigma^z_{ij}\sigma^z_{jk}\sigma^z_{kl}\sigma^z_{li}
	-\Gamma\sum_{\langle ij\rangle} \sigma^x_{ij} \nonumber \\
	&&-t\sum_{\langle ij\rangle,\sigma=\uparrow,\downarrow}
	\left( b^\dagger_{i\sigma}\; \sigma^z_{ij} \;b_{j\sigma} +{\rm H.c}\right)
	+\Delta \sum_{i \sigma} b^\dagger_{i\sigma} b_{i\sigma}\nonumber \\
	&&+V \sum_i \left[
	\left( b^\dagger_{i\uparrow} b_{i\uparrow} + b^\dagger_{i\downarrow} b_{i\downarrow} -\frac{1}{2}\right)^2
	-\frac{1}{4}\right]
	\label{eq:HK}
\end{eqnarray}
The operator $\sigma^z_{ij}$ has eigenvalues $\pm 1$, like a pseudo spin-$\frac{1}{2}$, and corresponds
to $e^{i\mathcal{A}_{ij}}$ in the path integral formulation (Eq.~\ref{eq:A}).
$\sigma^x_{ij}$ corresponds to the electric field operator.
In the path integral, $\mathcal{A}_{ij}$ and $\mathcal{E}_{ij}$ are conjugated.
So $\sigma^x_{ij}$ and $\sigma^z_{ij}$ should not commute on the same bond.
The natural choice in our discrete case is $\sigma^x_{ij}\sigma^z_{ij}=-\sigma^z_{ij}\sigma^x_{ij}$.
So, $\sigma^x_{ij}$ and $\sigma^z_{ij}$ are the $x$ and $z$ components of the pseudo spin-$\frac{1}{2}$.
The bosons represent the Bogoliubov quasi particles (spinon) of the mean field Hamiltonian.
The first term ($K$) is a sum over all the elementary plaquettes (square here for simplicity) and corresponds to the magnetic energy of the gauge field. The second term ($\Gamma$) is the electric energy, which generates fluctuations in the magnetic flux. The third one ($t$) describes the spinon hopping and their interaction with  the gauge field.
The last terms represents the energy cost $\Delta>0$ to create a spinon (related to the spin gap of the spin model)
and some (large) penalty $V$ when more than one spinon are on the same site.

This model is of course not directly  related to the original spin model but
contains the same two important ingredients that have been identified in the large $\mathcal{N}$ limit (spinon
coupled to $\mathbb Z_2$ gauge field fluctuations) and can provide as a simplified and phenomenological 
description to a gapped QSL. 

Because of the gauge symmetry, the physical Hilbert space of the model should be constrained to avoid
spurious degrees of freedom:
two states which differ by a gauge transformation correspond to a single physical state
and should not appear twice in the spectrum.
In the Hamiltonian formulation of gauge theories, the solution
is to construct the operators $U_{i_0}$ which generate the local gauge transformations, and impose
that all the physical states should be invariant under these transformations:
$U_{i_0}|{\rm phys.}\rangle=|{\rm phys.}\rangle\;\forall i_0$.
In the present case, an elementary gauge transformation
at site $i_0$ changes the value $\sigma^z_{i_0j}$
for all neighbors $j$ of $i_0$ (noted $j\in +$). In addition, it changes the sign of the boson operators in $i_0$.
This transformation is implemented by the following unitary operator
\begin{equation}
 	U_{i_0}=\exp\left[i\pi (
		b^\dagger_{i_0\uparrow}b_{i_0\uparrow} + b^\dagger_{i_0\downarrow}b_{i_0\downarrow}
	)\right]
	\prod_{j\in +} \sigma^x_{i_0j} 
	\label{eq:Uio}
\end{equation}
The constraint $U_{i_0}=1$ is the lattice version
of the Gauss law, ${\rm div} \vec E=\rho$, in electromagnetism, and the spinons
appear to play the role of the ``electric'' charges.

Readers familiar with lattice gauge theories will have recognized the Hamiltonian formulation of a $\mathbb Z_2$ gauge theory \cite{kogut79}. However, to show that the ground state of this model
realizes a topological phase (when $\Gamma$ is small enough), 
we will show that it is very close to the toric code model introduced by Kitaev \cite{kitaev97}.

\subsection{Toric code limit}

One goal of these notes was to show that (gapped) QSL in Mott insulators are
topologically ordered states with emerging gauge degrees of freedom.
To conclude, we will now take advantage of Kitaev's lectures of topological states of matter (in this school),  and show the close connection between the large $\mathcal{N}$ description of gapped QSL and Kitaev's toric code \cite{kitaev97}.

We consider the limit of Eq.~\ref{eq:HK}
when $t=0$, $\Gamma=0$ and $V=\infty$.
In this limit, the bosons cannot hop any more, and can only be zero or one per site:
$n_i=b^\dagger_{i_0\uparrow}b_{i_0\uparrow} + b^\dagger_{i_0\downarrow}b_{i_0\downarrow}\in\{0,1\}$.
Using  $U_{i}=1$ (Eq.~\ref{eq:Uio}) we find:
$e^{i\pi n_i}=\prod_{j\in +} \sigma^x_{ij}$,
so that the boson occupation numbers are  expressed
in terms of the (lattice divergence of the) electric field operators:
$2n_i=1-\prod_{j\in +} \sigma^x_{ij}$.
Taking 
The Hamiltonian can then be written as 
\begin{eqnarray}
 H&=&-K\sum_{\Box} \sigma^z_{ij}\sigma^z_{jk}\sigma^z_{kl}\sigma^z_{li}
	-\frac{1}{2}\Delta \sum_{i}	\prod_{j\in +} \sigma^x_{ij}
	\label{eq:HT}
\end{eqnarray}
which is {\it exactly} the (solvable) toric code Hamiltonian \cite{kitaev97}.

We can now import some  results from the toric code analysis.
Although simple derive in the framework of Eq.~\ref{eq:HT}, they are highly non trivial
from the point of view of the original spin model.
First, the ground state breaks no symmetry and 
the spinons (here at the sites $i$ with $\prod_{j\in +} \sigma^x_{ij}=-1$)
are free particles, they are not confined by the gauge field fluctuations.
Secondly, the ground state is degenerate on a cylinder or on a torus (periodic boundary conditions), as required by the LSMH theorem. The ground state are topologically ordered in the sense that no local observable can distinguish the different ground states.
Beyond the spinons, the model also have $\mathbb Z_2$-vortex excitations, which correspond
to plaquettes with $\sigma^z_{ij}\sigma^z_{jk}\sigma^z_{kl}\sigma^z_{li}=-1$. These gapped excitations are singlet states in the original spin model since the bond field $Q_{ij}$ and its sign fluctuations $\sigma^z_{ij}$ are rotationally invariant.\footnote{Ref.~\cite{bfg02} is an example of spin-$\frac{1}{2}$ model (with $U(1)$ symmetry)
where such a $\mathbb Z_2$ QSL is realized and where these vortex excitations, dubbed {\it visons}, have been studied. Visons have also bee  studied in the context of quantum dimer models, which are
simplified models for the the short range VB dynamics in frustrated quantum antiferromagnets
(see \cite{ml05} for a brief introduction).}
These excitation have a non trivial mutual statistics with respect to the spinon, and a bound state of a spinon and a vison behaves as a fermion.
Finally, the topological properties of the model (fractional excitations, topological degeneracy) are robust to perturbations, and should persist in presence of a small $\Gamma$ and small $t$ (Eq.~\ref{eq:HK}).


\begin{thebibliography}{99}

\bibitem{lsm61}
E. H. Lieb, T.D. Schultz, and D.C. Mattis,
\journaldoi{Ann. Phys. (N.Y.)}{16}{407}{1961}{10.1016/0003-4916(61)90115-4}.


\bibitem{hastings04}
M. Hastings, \journal{\prb}{69}{104431}{2004}.

\bibitem{kitaev97}
A.  Kitaev, \journaldoi{Ann. Phys.}{303}{2}{2003}{10.1016/S0003-4916(02)00018-0}.

\bibitem{taniguchi95}
S.~Taniguchi {\it et~al.},
\journal{\jpsj}{64}{2758}{1995}.

\bibitem{k99}
H. Kageyama {\it et~al.},
\journal{\prl}{82}{3168}{1999}.
\bibitem{tlcucl3}
K. Takatsu, W. Shiramura and H. Tanaka,
\journal{\jpsj}{66}{1611}{1997}; 
W.~Shiramura {\it et al.}, 
\journal{\jpsj}{66}{1900}{1997}.


\bibitem{herbert}
P. P. Shores {\it et al.}, 
\journal{J. Am. Chem. Soc.}{127}{13462}{2005};
J. S. Helton {\it et al.}, 
\journal{\prl}{98}{107204}{2007};
P. Mendels {\it et al.}, 
\journal{\prl}{98}{077204}{2007};
A. Olariu {\it et al.}, 
\journal{\prl}{100}{087202} {2008};
T. Imai {\it et al.}, 
\journal{\prl}{100}{077203}{2008}.

\bibitem{volbo}
Z. Hiroi {\it et al.},
\journal{\jpsj}{70}{3377}{2001};
F. Bert {\it et al.}, \journal{\prl}{95}{087203}{2005} 

\bibitem{shimizu03}
Y. Shimizu {\it et al.}, \journal{\prl}{91}{107001}{2003}.
 
\bibitem{itou08}
T. Itou {\it et al.}, \journal{\prb}{77}{104413}{2008} 

\bibitem{masutomi04}
M. Ryuichi, K. Yoshitomo  and I. Hidehiko, \journal{\prl}{92}{025301}{2004}


\bibitem{gaplessQSL}
W. Rantner and X.-G. Wen,
\journal{\prl}{86}{3871}{2001};
M. Hermele {\it et al.}, 
\journal{\prb}{70}{214437}{2004};
Y.~Ran, {\it et al.}, \journal{\prl}{98}{117205}{2007}
J. Alicea {\it et al}, 
\journal{\prl}{95}{241203}{2005}.

\bibitem{anderson52}
P. W. Anderson, \journal{\pr}{86}{694}{1952}.

\bibitem{hp40}
T. Holstein and H. Primakoff, \journal{\prb}{58}{1098}{1940}.

\bibitem{chs92}
J. T. Chalker, P. C. Holdsworth, and E. F. Shender,
\journal{\prl}{68}{855}{1992}.

\bibitem{cd88}
P. Chandra and B. Douçot, \journal{\prb}{38}{9335}{1988}.

\bibitem{arnold}
V. I. Arnold, {\it Mathematical Methods of Classical Mechanics}, Graduate Texts in Mathematics, Springer-Verlag, 1989.


\bibitem{affleck88}
I. Affleck,
\journal{\prb}{37}{5186}{1988}.

\bibitem{oya97}
M. Oshikawa, M. Yamanaka, and I. Affleck,
\journal{\prl}{78}{1984}{1997}.

\bibitem{oshikawa00}
M. Oshikawa,
\journal{\prl}{84}{1535}{2000}.

\bibitem{ns07}
 B. Nachtergaele and R. Sims,
\journaldoi{Com. Math. Phys.}{276}{437}{2007}{10.1007/s00220-007-0342-z}.

\bibitem{bfg02}
L. Balents, M. P. A. Fisher and S. M. Girvin,
\journal{\prb}{65}{224412}{2002};
D. N. Sheng and L. Balents,
\journal{\prl}{94}{146805}{2005}.

\bibitem{wen89}
X.-G. Wen, \journal{\prb}{40}{7387}{1989}.

\bibitem{wen91}
X.-G. Wen, \journal{\prb}{44}{2664}{1991}.

\bibitem{wn90}
X.-G. Wen and Q. Niu, \journal{\prb}{41}{9377}{1990}.

\bibitem{laughlin81}
R. B. Laughlin, \journal{\prb}{23}{5632}{1981}.

\bibitem{mg69}
C.~K.~Majumdar and D.~K.~Ghosh,
\journal{J. Math. Phys.}{10}{1399}{1969}.

\bibitem{klein82}
D. J. Klein,
\journaldoi{J. Phys. A: Math. Gen.}{15}{661}{1982}{10.1088/0305-4470/15/2/032}.

\bibitem{rms05}
K. S. Raman, R. Moessner and S.~L.~Sondhi,
\journal{\prb}{74}{064413}{2005}


\bibitem{cd94}
P. Chandra and B. Douçot,
\journaldoi{J. Phys. A: Math. Gen.}{27}{1541}{1994}{10.1088/0305-4470/27/5/019}


\bibitem{ss81}
B. S. Shastry and B. Sutherland,
\journaldoi{Physica (Amsterdam)}{108B}{1069}{1981}{10.1016/0378-4363(81)90838-X}.


\bibitem{anderson73}
P. W.~Anderson,
\journaldoi{Mat. Res. Bull.}{8}{153}{1973}{10.1016/0025-5408(73)90167-0}.

\bibitem{ml05}
G. Misguich and C. Lhuillier, in {\it Frustrated spin systems}
edited by H.~T.~Diep (World-Scientific, Singapore 2005)
(also on  \journal{cond-mat}{0310405}{}{}).

\bibitem{glkm07}
A.~Gell\'e {\it et al.}, 
\journal{\prb}{77}{014419}{2008}.

\bibitem{aa88}
D. Arovas and A.~Auerbach,
\journal{\prb}{38}{316}{1988}.

\bibitem{auerbachbook}
A. Auerbach, {\em Interacting electrons and Quantum Magnetism},
Springer-Verlag, 1994.

\bibitem{a85}
I. Affleck, \journal{\prl}{54}{966}{1985}

\bibitem{am88}
I. Affleck and J. B. Marston,
\journal{\prb}{37}{3774}{1988};
J. B. Marston and I. Affleck,
\journal{\prb}{39}{11538}{1989}


\bibitem{rs89}
N. Read and S. Sachdev,
\journal{\prl}{62}{1694}{1989}.

\bibitem{rs91}
N. Read and S. Sachdev,
\journal{\prl}{66}{1773}{1991}.

\bibitem{wen02}
X.-G. Wen,
\journal{\prb}{65}{165113}{2002}.



\bibitem{kogut79}
J. B. Kogut,
\journal{\rmp}{51}{659}{1979}.


\end{thebibliography}
\end{document}